# Dipolar excitations at the L$_{III}$ x-ray absorption edges of the heavy rare earth metals

S. D. Brown,[1,8] P. Strange,[2] L. Bouchenoire,[1,8] B. Zarychta,[3] P. B. J. Thompson,[1,8] D. Mannix,[1,8] S.J. Stockton,[3] M. Horne,[3] E. Arola,[3,4] H. Ebert,[5] Z. Szotek,[6] W. M. Temmerman,[6] and D. Fort[7].

[1] XMaS, European Synchrotron Radiation Facility, B.P. 220, 38043 Grenoble Cedex, France
[2] School of Physical Sciences, University of Kent, Canterbury, Kent, CT2 7NH, UK.
[3] Keele University, Keele, Staffordshire ST5 5BG, UK.
[4] Optoelectronics Research Centre, Tampere University of Technology, P.O. Box 692, FIN-33101, Tampere Finland.
[5] Department Chemie/Physikalische Chemie,Universit*a*t Munchen, butenandtstrasse 5-13,D-81377, Munich, Germany.
[6] Daresbury Laboratory, Daresbury, Warrington, WA4 4AD, UK.
[7] The University of Birmingham, Edgbaston, Birmingham B15 2TT, UK
[8] Liverpool University, Liverpool L69 3BX, UK
(Dated: December 5, 2006)

We report measured dipolar asymmetry ratios at the L$_{III}$ edges of the heavy rare earth metals. The results are compared with a first principles calculation and excellent agreement is found. A simple model of the scattering is developed, enabling us to re-interpret the resonant x-ray scattering in these materials and to identify the peaks in the asymmetry ratios with features in the spin and orbital moment densities.

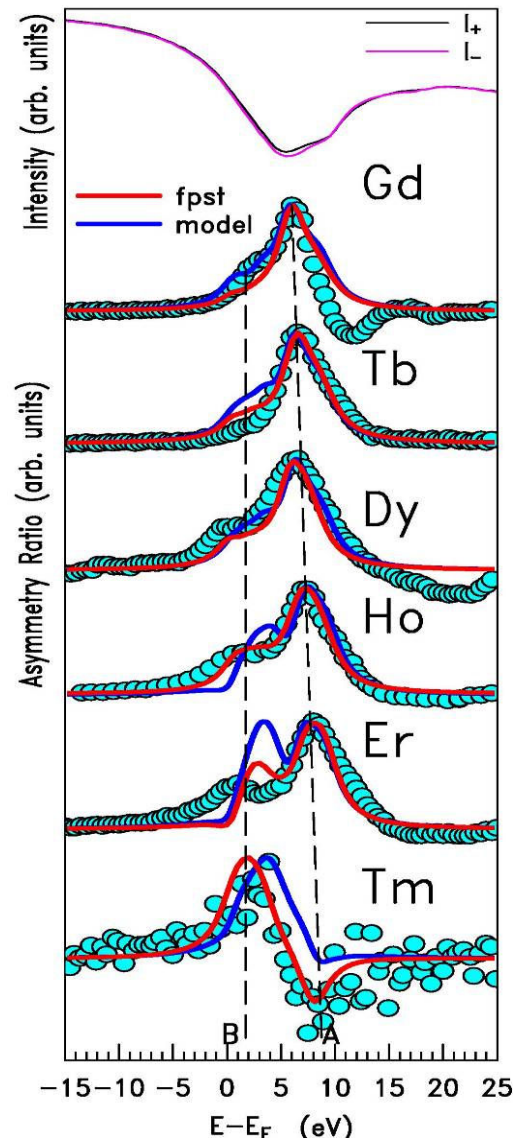

The interpretation of magnetic x-ray spectroscopies at the $L$ absorption edges of rare earth materials has been a subject of long standing controversy. Spectra consist of two main features: a large peak at an energy close to the absorption white-line (feature A), interpreted as arising from dipolar ($E1$) transitions coupling $2p$ core levels to unoccupied $5d$ and $6s$ states; a smaller peak a few eV lower (feature B), attributed to quadrupolar ($E2$) transitions coupling the same core levels to unoccupied $4f$ states [1-6]. In this paper we re-examine the dipolar/quadrupolar character of these structures.

X-ray resonant magnetic scattering (XRMS) assigns the $E1$ or $E2$ nature of spectral features through analysis of the scattered x-ray polarisation [7,8] and comparison with theory [9-11]. Resonant inelastic x-ray scattering (RIXS) assigns the $E1$ or $E2$ nature from their relative energy positions [5,6]. In x-ray magnetic circular dichroism (XMCD) assignment is achieved by modelling variations in signal with the angle between the incident x-ray polarisation and the magnetisation direction [12]. However, problems arise in the sign of signals and the relative importance of $E2$ 'contamination'. Spectra show dispersive shapes which are thought to arise from exchange splitting and matrix element effects [13,14].

In this letter we report combined experimental and theoretical investigations of x-ray resonant interference scattering (XRIS) in the ferro/ferrimagnetic phases of heavy rare earths. For the heavy rare earth hexagonal crystal structures, $E1$-$E2$/$M1$ scattering is allowed and may contribute to asymmetry ratios. We exploit the possibility of virtually turning off the $E2$ signal by scattering horizontally through angles close to 90°, thus isolating $E1$ scattering. To investigate this further, first principles calculations were performed for both pure $E1$, pure $E2$, and total ($E1$ and $E2$) scattering in this geometry. The $E1$-$E2$ contributions were found to be two orders of magnitude smaller than the pure $E1$ scattering and thus can be eliminated as key contributions. We show that feature B is in fact of mixed $E1$ and $E2$ origin in both absorption and scattering spectra and that the $E1$ contribution is substantial. Using both first principles scattering theory (fpst) and an atomic-like model, we reproduce features A and B and thus confirm the $E1$ nature of our spectra. This has enabled identification of strong antiparallel hybridisation between the unoccupied $4f$ and $5d$ spin moments which provides a natural explanation for the size and dispersive line-shape observed in XMCD. A direct hybridisation between $f$ and $d$ states with parallel orbital moments is also identified. A changeover from magnetism dominated by $f$-$d$ exchange and spin, to magnetism dominated by spin-orbit coupling (SOC) and orbital $4f$ and $5d$ magnetisations occurs around Er. However, with both the fpst and the model, the Tm XRIS spectra were found to be inverted with respect to the experimental data. Examining this discrepancy suggests the possibility that in the excited state, the $4f$ SOC might differ from that of the Hund's rules ground state configuration, although we stress that assignment of an $E1$ contribution to feature B is independent of this possibility.

**Fig 1.** Top curves; field dependent energy scan for Gd. At each energy point the field was reversed many times and averaged to reduce systemativ errors. Below; charge-magnetic XRIS asymmetry ratios (see equation (3)). . Experimental data (cyan), first-principles x-ray scattering theory (red) and model (blue). Statistical errors are smaller than the plotted points In Tm, (with scattering angle < 90 °), the asymmetry ratio reverses sign. Therefore we have simply inverted the curves for Tm in the figure. The curves consist of a dominant high energy peak and a lower energy shoulder that becomes more pronounced with increasing atomic number. The higher energy peak results from scattering off the spin moment and the lower energy peak arises from scattering from both the spin and orbital moments. The changing relative size and position of these peaks reflects the change in relative magnitude of these quantities. The zero of energy here is the $L_{III}$ edge which we equate to the Fermi energy.

The experiments were performed at the XMaS beamline BM28 at the European Synchrotron Radiation Facility [15]. Ferri/ferromagnetic charge-magnetic XRIS measurements were performed monitoring (300) reflections from each element, whilst scattering horizontally through angles of about 90° at their $L_{III}$ edges. Data were obtained through reversal of a vertical magnetic field applied along the $c$-axis at each point of energy scans, resulting in field dependent spectra, an example of which is given in the top pannel of Fig. 1 for Gd. It is usual to present such data as an asymmetry ratio, $R$, as defined in eq. 3. [16]. . Data were obtained through reversal of a vertical magnetic field applied along the $c$-axis Spectra were calculated from a fully relativistic fpst of XRMS [17,18]. The electronic structure of the elements was found using the fully relativistic SIC-LMTO method [19,20].

To model the XRIS spectra, we write the total $E1$ scattering amplitude as a sum of amplitudes due to charge scattering (subscript 0), spin (subscript S) and orbital (subscript L) magnetic scattering:

$$f_{E1}(E) = -\left(\frac{3}{4\pi q}\right)\{\cos 2\theta (F_0' + iF_0'') + i\mathbf{m}\sin 2\theta (F_S' + iF_S'' + F_L' + iF_L'')\} \quad (1)$$

Where single (double) primes indicate the real (imaginary) part of the resonant scattering amplitudes $F$, and **m** is a unit vector in the magnetisation direction. On reversal of the magnetisation, magnetic terms in equation (1) change sign while the charge terms do not. The absorption coefficient is directly proportional to the DOS. In turn the Optical Theorem [21] states that the absorption coefficient is directly proportional to the imaginary part of the forward scattering amplitude. We therfore replace the imaginary parts of the scattering amplitudes of equation (1) with the equivalent DOS found from the electronic structure calculations and their real parts with a Kramers-Kronig transformation (KKT)[22] of these quantities. This method ignores matrix element effects which we have found to be slowly varying over the energy range of interest. It also implies that the extra transparency this model yields will lead to only a small accuracy loss. We define [10];

$$\hat{\rho}_0 = \hat{\rho}_\uparrow + \hat{\rho}_\downarrow, \text{ and } \hat{\rho}_S = \frac{1}{2}(\hat{\rho}_\uparrow - \hat{\rho}_\downarrow) \quad (2)$$

as the charge and spin polarised convoluted densities of states respectively, where $\hat{\rho}\uparrow(\downarrow)$ is the spin up (down) unoccupied DOS convoluted with a Lorentzian (here, and in the fpst, of width 3 eV for all elements) representing the core hole lifetime. Replacing the scattering amplitudes in equation (1) with the appropriate unoccupied DOS we arrive at;

$$R = \frac{I_+ - I_-}{I_+ + I_-} = \frac{2\tan 2\theta \{2P_S(\hat{\rho}_\downarrow \hat{\rho}_\uparrow'' - \hat{\rho}_\uparrow \hat{\rho}_\downarrow'') + P_L(\hat{\rho}_0 \hat{\rho}_L'' - \hat{\rho}_0'' \hat{\rho}_L)\}}{\hat{\rho}_0''^2 + \hat{\rho}_0^2} \quad (3)$$

where, $I_+$ is the cross section for the magnetisation in one direction and $I_-$ is for it reversed, the $\hat{\rho}''$'s represent the KKT's of the DOS. $P_S$ and $P_L$ are the same photoelectron spin and orbital polarisations adopted in the analysis of XMCD and we have taken the simple values $|P_S|= 1/4$ and $|P_L|= 3/4$. Our theoretical and experimental approach of measuring photon scattering has significant advantage over first order methods (absorption, emission): (i) the XRIS asymmetry ratio is enhanced by a factor of $2\tan(2\theta)$, leading to easily measurable signals; (ii) Diffraction is a highly efficient scattering mechanism leading to enhancement of the signal to noise ratio; (iii) we reference the signs of both the spin and the orbital contributions (and their real parts) to those of the charge, which enhances sensitivity; (iv) polarisation dependence allows almost complete separation of $E1$ and $E2$ events. Finally, in contrast to antiferromagnetic XRMS, no absorption correction is required.

Fig. 1 shows measured XRIS asymmetry ratios at the $L_{III}$ edges. The spectra resemble RIXS spectra in [6], where the low energy feature was designated as $E2$ from its energy position. Also shown are pure $E1$ results of the fpst and model. The experimental results have been rigidly shifted to superpose the high energy peaks with those in the calculated spectra and rescaled. In all elements the data contain an $E1$ peak, feature A, with a lower energy $E1$ shoulder/peak, feature B. For Tm, Feature A has become inverted, moreover, for the Hund's rule ground state $4f$ electronic configuration (J=L+S), both the fpst and the model give a negative peak at low energy and a positive peak at high energy. Er also shows a dispersive line-shape for the J=L+S 4*f* configuration. Small improvements in agreement between experiment and both calculations are observed for the J=|L-S| 4*f* configuration for all elements. Therefore we show the non-Hund's rules compliant results, although we note that for Gd to Ho this makes very little difference and that even for Er and Tm, features A and B still occur at the same energies as those observed with the Hund's rules configuration.

The *E*2 scattering amplitude simplifies in the adopted geometry, just one reversible magnetic term remains. The total *E*1+*E*2 asymmetry ratio leads to pure *E*1, pure *E*2 and mixed *E*1*E*2 charge magnetic interference terms. *E*2 charge scattering events are two orders of magnitude smaller than *E*1 charge scattering events. However, this is not true for *E*2 magnetic scattering, which is enhanced relative to *E*1 magnetic scattering, due to the large 4*f* and small 5*d* moments. This yields asymmetry ratios where the *E*1 charge-*E*2 magnetic term is scaled by a factor $\cos(2\theta)$ relative to pure *E*1 terms. Thus for Ho, Er and Tm ($2\theta$ =96.1º, 93.3º and 89.5º) the *E*2 contribution are reduced by factors of 0.10, 0.06 and 0.01 respectively. For Gd, Tb, and Dy the scattering angles exceed 100º and the *E*2 suppression is less efficient. Clearly both the fpst and the model predict *E*1 scattering at energy position B for these elements, which fits the experimental data. We have performed polarisation analysis in the antiferromagnetic phases of Tb to Tm, which reveals almost pure E1 scattering in Ho, Er and Tm and sets an upper limit of the *E*2 contribution in Tb and Dy of a few percent [16].

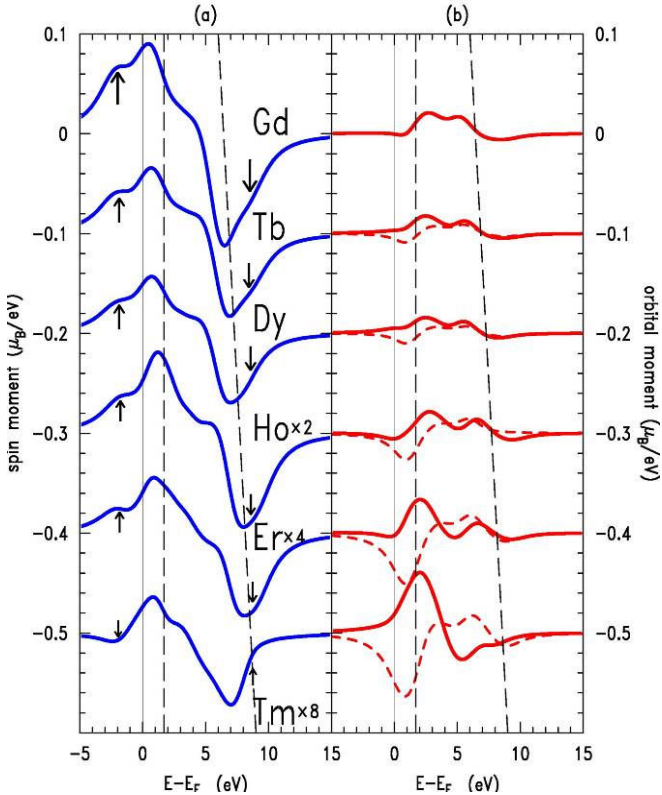

Fig 2. (a) Spin polarised convoluted 5*d* DOS: Beyond Gd, the data are offset by -0.1 $\mu_B$/eV with respect to the preceding element. The DOS have been convoluted with a 3 eV wide Lorentzians, to represent core hole broadening. The centre of the occupied 5*d* spin moments is also shown below $E_F$, indicated by the arrows. The arrows above the $E_f$ show the energy of the centre of the 5*d* spin moment closure peaks. The near vertical dashed lines mark the position of features A and B in Fig 1. (b) Orbital polarised convoluted 5*d* DOS: Data are scaled and offset as in (a). The dashed line represents the Hund's rule state and the full line is for the non-Hund's rule state.

The Hund's rules ground state calculations give total magnetic moments in fair agreement with measured values for Gd to Tm. Exchange splitting dominates for Gd to Ho, after which SOC leads to bunching of mixed spin up/spin down states around the energies of the Lu j=5/2 and 7/2 closed shell configuration. This is seen in the exchange splitting between occupied 5*d* spin states, which decreases form 0.63 eV in Gd to 0.04 eV in Tm. The occupied 5*d* spin moment (0.292 $\mu_B$ in Gd, 0.023 $\mu_B$ in Er and -0.026 $\mu_B$ in Tm) is aligned with that of the 4*f*'s for Gd to Er but is antiparallel for Tm. They exhibit dispersive line-shapes, as seen in Fig 2(a). The positive parts of the dispersive shapes are stationary at around 1 eV above the Fermi energy for all elements. These features coincide with negative peaks in the unoccupied 4*f* spin moments, shifted to slightly higher energy due to the large increase in available *d*-states from 0 to 1 eV. The negative parts of the dispersive shapes occur at higher energies; ~6.5 eV for Gd, increasing to ~8 eV for Er, reducing to 7 eV for Tm. The stationary nature of the low energy peaks leads us to conclude that they arise from antiparallel *f-d* spin hybridisation. This large positive contribution to the 5d unoccupied moment must be compensated and this is the origin of the large negative part of the dispersive oscillation at high energy. In Tm, where the occupied *d* moment is reversed relative to that of the 4*f* states, this feature changes in a positive sense, as indicated by the arrows in Fig 2(a). This is also reflected in the polarity of feature A in Fig 1. We also find that the 6*s* unoccupied spin moment does not exhibit the same dispersive shape as the 5*d*'s. The parts of the unoccupied 6*s* spin states which negatively reflect the occupied moments, are hybridised with the 5*d* DOS in the high energy tails of the 5*d* band. Clearly, this large 4*f* induced oscillation must be the origin of the dispersive XMCD spectra. Even if there were no *d*-band polarisation, this dispersive feature would be present, though its integral would be zero. In common with Kim *et. al.*[14] we observe an increase in the size of the matrix elements as energy increases.

The 5*d* orbital moments are shown in Fig. 2(b). They exhibit double peaks, split by about 4 eV, for all elements except Tm. Here, the centroids of the two peaks are aligned with the centres of the 5*d* bands, which are at 3.5 to 5 eV for Gd to Tm. This splitting is visible in the total 5*d* DOS and is also apparent in the unoccupied spin moment in Fig.2 (a). We conclude that this splitting is due to the crystal field (CEF) (which is essentially unaffected by the partially filled f-shell configuration). In addition, there are negative dips in the Hund's rules spectra (dashed lines) which are also stationary, at around 1 eV above $E_f$ for all elements. These features coincide with the negative peaks in the unoccupied 4*f* orbital moments, again, slightly shifted up in energy due to available *d*-states. Further, if we reverse the sign of the 4*f* spin orbit coupling (non-Hund's rules configuration; solid lines), these features change sign and are pulled down in energy along with the unoccupied 4*f* orbital moments. We conclude that these features are due to parallel *f-d* orbital hybridisation. The higher energy 5*d* orbital moments are fairly independent of the 4*f* orbital configuration, as was concluded in [14]. For Er and Tm however, the hybridised 4*f* orbital moments dominate over the CEF splitting. Our results strongly suggest that in the presence of the $2p^{3/2}$ core hole, the 4*f* SOC differs from that of the Hund's rules ground state. It has been suggested [23] that in the presence of the core hole, the 4*f* levels pull down in energy by ~ 3 eV (although earlier calculations found larger values[24]). As this also corresponds to calculated energy difference between the J=L+S and |L-S| 4*f* configurations, it is conceivable that the core hole energy might be taken up in a reconfiguration of the 4*f* states, rather than a rigid shift of them during the excitation and our results support this scenario. Examination of the positions of features A and B in Fig. 2 reveals that feature A is of almost pure spin origin whereas feature B is of mixed spin and orbital origin. The correlation of peaks A and B in Fig. 1, with structure in Fig. 2(a) and Fig. 2(b) demonstrates that the dipolar asymmetry ratios at the $L_{III}$ edges reflect the progression from spin-dominated Gd to orbital-dominated Er and Tm electronic structure.

This work, with analogous work at the $L_{II}$ edge, opens the way for quantitative measurements of quadrupole scattering contributions and the development of new sum rules.

We gratefully acknowledge the financial support of EPSRC. The experimental work was performed on the EPSRC-funded XMaS beamline at the ESRF, directed by M. J. Cooper and C. A. Lucas. SDB would like to thank S. P. Collins and S. W. Lovesey and staff of the ESRF, notably C. Detlefs, for stimulating discussions.

[1] G. Shutz, W. Wagner, W. Wilhelm, P. Kienle , R. Zeller, R. Frahm and G. Materlik , Phys. Rev. Lett. **58**, 737 (1987).
[2] P. Fischer, G. Schutz., S. Stahler and G. Weisinger, J. Appl.Phys. **69**, 6144 (1991)
[3] F. Baudelet, C. Giorgetti, S. Pizzini, C. H. Brouder, E. Dartyge, A. Fontaine, J. P. Kappler and G. Krill, J. Electron Spectrosc. Relat. Phenom. **62**, 153 (1993)
[4] D. Gibbs, D. R. Harshman, E. D. Isaacs, D. B. McWhan, D. Mills and C. Vettier, Phys. Rev. Lett. **61**, 1241 (1988)
[5] M. H. Krisch, C. C. Kao, F. Sette, W. A. Caliebe, K. Hamalainen and J. B. Hastings, Phys. Rev. Lett. **74**, 4931 (1995).
[6] F. Bartolome, J. M. Tonnerre, L. Seve, D. Raoux, J. Chaboy, L. M. Garcia, M. Krisch and C. C. Kao, Phys. Rev. Lett. **79**, 3775 (1997).
[7] D. Gibbs, G. Grubel, D. R. Harshman, E. D. Isaacs, D. B. McWhan, D. Mills and C. Vettier, Phys. Rev. B **43**, 5663 (1991)
[8] C. Detlefs, A. H. M. Z. Islam, A. I. Goldman, C. Stassis, P. C. Cranfield, J. P. Hill and D. Gibbs, Phys. Rev. B **55**, R680 (1997)
[9] M. Blume, J. Appl. Phys. **57**, 3615 (1985)
[10] S. W. Lovesey, and S. P. Collins, *X-ray Scattering and Absorption by Magnetic Materials,* (Oxford University Press, Oxford 1996).
[11] J. P. Hill and D. F. McMorrow, Acta Cryst. A**52**, 236 (1996).
[12] J. C. Lang, G. Srajer, C. Detlefs, A. I. Goldman, H. Konig, Wang Xindong, B. N. Harmon and R. W. McCallum, Phys. Rev. Lett. **74**, 4935 (1995)
[13] M. van Veenendaal, J. B. Goedkoop and B. T. Thole, Phys. Rev. Lett. **78**, 1162 (1997)
[14] J. W. Kim, Y. Lee, D. Wermeille, B. Sieve, L. Tan, S. L. Bud'ko, S. Law, P. C. Cranfield, B. N. Harmon and A. I. Goldman, Phys. Rev. B **72**, 064403 (2005)
[15] S. D. Brown, L. Bouchenoire, D. Bowyer, J. Kervin, D. Laundy, M. J. Longfield, D. Mannix, D. F. Paul, A. Stunault, P. Thompson, M. J. Cooper, C. A. Lucas and W. G. Stirling, J. Synch. Rad. **8**, 1172 (2001).
[16] S. D. Brown, P. Strange, L. Bouchenoire, B. Zarychta, P. B. J. Thompson, D. Mannix, S.J. Stockton, M. Horne, E. Arola, H. Ebert, Z. Szotek, W. M. Temmerman, and D. Fort, to be submitted to Phys. Rev. B.
[17] E. Arola, P. Strange and B. L. Gyorffy, Phys. Rev. B **55** 472 (1997)
[18] E. Arola, M. Horne, P. Strange, H. Winter, Z. Szotek and W. M. Temmerman, Phys. Rev. B **70**, 235127 (2004).
[19] W. M. Temmerman, A. Svane, Z. Szotek and H. Winter, *Electronic Density Functional Theory: Recent Progress and New Directions*, edited by J. F. Dobson, G. Vignale and M. P. Das (Plenum 1997).
[20] S. V. Beiden, W. M. Temmerman, Z. Szotek and G. A. Gehring, Phys. Rev. Lett. **79,** 3970
[21] G. Baym, *Lectures on Quantum Mechanics* (Addison-Wesley, California, 1990)
[22] C. Kittel, *Introduction to Solid State Physics* (John Wiley and Sons, New York 1986)
[23] P. Carra, B. N. Harmon, B. T. Thole, M. Altarelli and G. A. Sawatzky, Phys. Rev. Lett. **66**, 2495 (1991)

[24] J. F. Herbst and J. W. Wilkins, *Handbook of the Physics and Chemistry of the Rare Earths* **10**, 356, (1987) and references therein.